\documentclass[twocolumn,showpacs,aps,prd]{revtex4}

\usepackage{graphicx}
\usepackage{dcolumn}   
\usepackage{amsmath} 
\usepackage{epsfig}
\usepackage{hhline}
\RequirePackage{xspace}
\usepackage{bm}

\def\sbar{{\overline s}}
\def\figurebox#1#2#3{%
    \def\arg{#3}%
    \ifx\arg\empty
    {\hfill\vbox{\hsize#2\hrule\hbox to #2{\vrule\hfill\vbox to #1{\hsize#2\vfill}\vrule}\hrule}\hfill}%
    \else
    {\hfill\epsfbox{#3}\hfill}%
    \fi}

\newcommand{\non}{\nonumber\\}

\begin{document}

\preprint{LBNL-xxxx}
\begin{flushleft}
LBNL-54065\\
\end{flushleft}

\title{
{\large \bf Experimental Limits on the Width of the Reported $\Theta(1540)^+$ 
}}

\author{ Robert N. Cahn and George H. Trilling}
\affiliation{Lawrence Berkeley National Laboratory\\ 1 Cyclotron Rd.,\\
Berkeley, CA 94720)}
\date{\today}

\begin{abstract}
Using data on $K^+$ collisions on xenon and deuterium we derive values
and limits on the width of the reported $\Theta(1540)^+$ exotic baryon
resonance.  The xenon experiment gives a width of $0.9\pm 0.3$ MeV.  The 
other experiments give upper limits in the range 1 - 4 MeV.

\end{abstract}

\pacs{14.20.-c, 13.75.-n}
\maketitle

\newpage

\setcounter{footnote}{0}
\section{\boldmath Introduction}

The general features of the spectroscopy of mesons and baryons can be
understood from the simple rules that a meson is made of a quark and
an antiquark, while a baryon is made from three quarks.  These rules
are consistent with the principles of quantum chromodynamics, which
show that physical particles are neutral under color-$SU(3)$.
However, QCD does not preclude the existence of other colorless
configurations, including gluons or additional quarks and antiquarks.
Recent results from a diverse collection of experiments \cite{LEPS,
  DIANA, CLASI, CLASII, SAPHIR, neutrino} show evidence for a state
$\Theta(1540)^+$, whose quantum numbers are those of the combination
$uudd\sbar$ and which thus cannot be composed simply of three quarks.
Such states have been predicted \cite{Diakonov} in a Skyrmion
model.  

The $\Theta(1540)^+$ has not been detected in data from a
number of early experiments in which one might expect it to appear.
However, definitive negative conclusions cannot be drawn in the
absence of reliable predictions for production cross sections.

In situations where the $\Theta(1540)^+$ is or ought to be formed
resonantly, as an intermediate state in a scattering experiment, it is
possible to draw conclusions about its width from existing data.  Such
results can provide guidance to the structure of the resonance, or
more important, to the likelihood that there truly is a such
resonance.

The resonant cross section is determined entirely by $\Gamma$, the
width of the resonance and its branching ratios $B_i$ and $B_f$ into
the initial and final channels according to the Breit-Wigner form:

\begin{equation}
\sigma(m)=B_i\, B_f\,\sigma_0\,\left[{\Gamma^2/4\over
    (m-m_0)^2+\Gamma^2/4}
\right]
\end{equation}
With $k$ as the CM momentum, $s_1$ and $s_2$ the incident spins
and
$J$ the spin of the resonance, we have
\begin{equation}
\sigma_0= {2J+1\over (2s_1+1)(2s_2+1)}{4\pi\over k^2}
= 68\ {\rm mb}
\end{equation}
where we have taken the values for $K^+n$ collisions at a resonant
mass $m_0$
of 1540 MeV and assumed that the resonance has $J=1/2$.
As we shall see, the mass resolution of the experiments is always
broader than the natural width of the resonance so that the observable
quantity is the integral of the resonant cross section:

\begin{eqnarray}
&&\int_{-\infty}^\infty dm\, B_i\,B_f\,\sigma_0\left[{\Gamma^2/4\over
    (m-m_0)^2+\Gamma^2/4}\right]\nonumber\\
&&\qquad\qquad=\, B_i\,B_f\,{\pi\Gamma\over
    2}\sigma_0\nonumber\nonumber\\
&&\qquad\qquad=(107\, {\rm mb})\times B_i\,B_f\, \Gamma\label{eq1}
\end{eqnarray}

\section{$K^+n\to K^0 p$ in xenon}
In the DIANA experiment \cite{DIANA}, in which a $K^+$ beam with momentum 750 MeV
entered
a xenon bubble chamber, the signal for the $\Theta(1540)^+$ was
observed by measuring the $p K_S$ invariant mass spectrum in the final state.  If we treat the
scattering as simply a two-body process, $K^+n\to K^0p$, resonance
occurs when the combination of the incident momentum of the $K^+$ and
the Fermi momentum of the neutron give the invariant mass of the
$\Theta(1540)^+$.  Without the Fermi momentum, this would occur for a $K^+$
momentum of 440 MeV, to which the incident beam is reduced by
ionization  losses after penetrating a sufficient distance through the
xenon.  By observing the final-state invariant mass, reconstructed
from
the $p K_S $, the effects of Fermi motion and incident beam
degradation are removed, provided that we can ignore rescattering within
the nucleus.

The signal in this experiment emerges only after making cuts that are
believed to reduce the effect of rescattering. We make the assumption
that, in the mass region near the resonance, it is the charge-exchange
process on a single nucleon
that is observed and that the cuts reduce the resonant and
non-resonant charge-exchange processes by the same factor.  The
apparent resonant signal is contained within two 5-MeV bins.  The
background varies smoothly in this region at a value near 22 events
per bin.  The resonant signal consists of about 26 events. We
associate the background events with the $K^+d$ charge exchange cross
section interpolated from off-resonance measurements, namely $4.1\pm 0.3$ mb
\cite{slater,damerell}.  In this way we determine the integral of the
resonant cross section to be $(26/22)\times 5\ {\rm MeV} \times 4.1\ 
{\rm mb}= 24\ {\rm mb\ MeV}$.  Using Eq.(\ref{eq1}), and $B_i=B_f=1/2$,
appropriate for either $I=0$ or $I=1$, we deduce a width
$\Gamma=0.9\pm 0.3\ {\rm MeV}$, where the quoted error is statistical
only.  There are systematic uncertainties, which we are unable to evaluate,
associated with rescattering in the nucleus and with the cuts that
isolate the signal. Of course, it is assumed that the excess events
are indeed the result of a resonance $\Theta(1540)^+$ and not an
artifact.

\section{$K^+d$}
If the decay products $K^+n$ or $K^0p$ are not measured precisely, the
collision energy with a nuclear target is necessarily uncertain as a
consequence of Fermi motion.  In the case of a deuterium target, the
Fermi
motion can be treated quite completely.  Let the momentum of the
incident $K^+$ be $P_K$ and let the component of the neutron's Fermi
momentum in the beam direction be $p_z$.  Then the CM energy squared
is
\begin{eqnarray}
s&=&m_K^2+2E_Km_N+m_N^2-2p_z P_K\nonumber\\
 &=&m_0^2+2(E_K-E_K^*)m_N-2p_zP_K
\end{eqnarray}
where we have introduced $E_K^*$ as the beam energy that would make
the resonance in the absence of Fermi motion.  Because the resonance
is narrow, we can write the Breit-Wigner form as
\begin{equation}
\sigma(E_K,p_z)=\sigma_0\, B_i\, B_f{\pi\Gamma\over 2}m_0\,
\delta((E_K-E_K^*)m_N-p_zP_K)
\end{equation}
   
The distribution $F(p_z)dp_z$ is related to the
full Fermi momentum distribution $f(p)d^3p$ by
\begin{equation}
F(p_z)=2\pi\int_{|p_z|}^\infty \,dp\, p f(p)
\end{equation}
In terms of $F(p_z)$, the resonant cross section, integrated over the
distribution of Fermi momenta is
\begin{eqnarray}
\sigma(E_K)&=&\sigma_0\, B_i\, B_f{\pi\Gamma\over 2}{m_0\over
  P_K}F((E_K-E_K^*)m_N/P_K)\nonumber\\
&=&(372 \,{\rm mb}) \times B_i\, B_f\, \Gamma\, F((E_K-E_K^*)m_N/P_K)\nonumber\\
\end{eqnarray}

The momentum-space wave function for the deuteron is easily computed
from a spatial wave function in the Hulth\'en form \cite{hulthen}:
\begin{eqnarray}
\phi(r)&=&N{1\over r}(e^{-\alpha r}-e^{-\beta r}) \nonumber\\
N^2&=&{\alpha\beta\over 2\pi}{\alpha+\beta\over (\alpha-\beta)^2}
\end{eqnarray}
Here $\alpha$ is related to the deuteron binding energy $\Delta E$ by
$\alpha=\sqrt{m_N\Delta E}=45.5\ {\rm MeV}$. A typical value for
$\gamma\equiv\beta/\alpha$ is 6.16. 
The momentum-space wave function is
\begin{eqnarray}
\tilde\phi(p)&=&\int d^3r {e^{-i{\bm p}\cdot{\bm r}}\over(2\pi)^{3/2}} \phi\non
&=&{\sqrt{\alpha\beta}\over\pi}{(\alpha+\beta)^{3/2}\over
  (\alpha^2+p^2)(\beta^2+p^2)}
\end{eqnarray}

\begin{figure}[h]
\includegraphics[width=2.1in,angle=90]{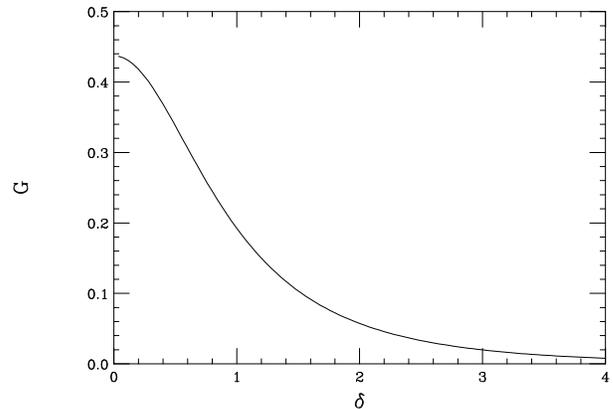}
\caption{The function $G(\gamma,\delta)$, which gives the density of 
nucleons in the deuteron as a function of momentum  in a given
direction $p_z=\alpha
\delta$, where $\alpha=45.5 \ {\rm MeV}$.\label{fig:1}}
\end{figure}

The momentum distribution is $f(p)=|\tilde\phi(p)|^2$. Setting
$F(p_z)=\alpha^{-1}G(\gamma,\delta)$, with $\delta=p_z/\alpha$, we find directly
\begin{eqnarray}
G(\gamma,\delta)&=&(1+\gamma)^3\gamma\left\{{1\over(\gamma^2-1)^2}\left[{1\over
          1+\delta^2}+{1\over \gamma^2+\delta^2}\right]\right.\nonumber\\
&&\left.-{2\over (\gamma^2-1)^3}\ln{\gamma^2+\delta^2\over 1+\delta^2}\right\}
\end{eqnarray}
The function $G(\gamma,\delta)$ is shown for $\gamma=6.16$ in Fig.~\ref{fig:1}.

Exactly at resonance we have $G(\gamma=6.16,\delta=0)=0.438$ and thus
$F(0)=0.0096\,{\rm MeV}^{-1}$. For a $K^+$ beam that is nearly tuned
to the resonant momentum of 440 MeV we then have for the cross section
averaged over Fermi momenta,
\begin{equation}
\sigma(E_K^*)=(3.6\, {\rm mb/MeV})\, B_i\, B_f\,  \Gamma
\end{equation}

Charge exchange in $K^+d$ collisions has been measured by Slater et
al.\cite{slater} at incident momenta of 376 MeV and 530 MeV and by
Damerell et al.\cite{damerell} at a momentum of 434 MeV.  From
Fig.~\ref{fig:1} a direct calculation shows that the former two
measurements are sufficiently far from the resonance to be unaffected,
while the last is so close that we can take $p_z=0$.  The measurements
are listed in Table~\ref{table:1}.

\begin{table}[h]
\begin{tabular}{ll}\hline 
$P_K$(MeV)&$\sigma_{CEX}$(mb) \\ \hline
376\ \cite{slater} & $3.1\pm 0.4$\\
434\ \cite{damerell} & $4.0\pm 0.15$\\
530\ \cite{slater} & $6.5\pm 0.6$ \\ \hline
\end{tabular}
\caption{Charge-exchange cross section in $K^+d$ reactions measured by
 Refs.~\cite{slater,damerell}.  \label{table:1}}
\end{table}

Comparison of the cross sections shows no sign of an enhancement near
440 MeV.  We take a conservative limit of 1 mb on the resonant cross
section.
Using $B_i=B_f=1/2$, we derive a limit of 1.1 MeV.

A similar analysis can be made using the total cross section data of
Bowen \cite{bowen}.  The measurements are shown in Table~\ref{table:2}.
Here linear interpolation shows an excess of $0.60\pm 0.30$ mb at the
resonant momentum, though this might well be simply a sign of a
gradual deviation from linearity.  We  take 1.5 mb as a conservative upper
limit
on the resonant cross section.  Then using $B_i=1/2$ and $B_f=1$, we
find an upper limit of $\Gamma=0.8$ MeV.

\begin{table}[h]
\begin{tabular}{ll}\hline 
$P_K$(MeV)&$\sigma_{tot}$(mb) \\ \hline
366 & $21.41\pm 0.30$\\
440 & $23.46\pm 0.24$\\
506 & $24.16\pm 0.23$ \\ \hline
\end{tabular}
\caption{Total  $K^+d$ cross section  measured by
 Ref.~\cite{bowen}.  \label{table:2}}
\end{table}

An even more conservative limit is obtained by taking the entire $I=0$
cross section at $P_K=440$ MeV to be resonant.  This cross section is
reported  by Bowen \cite{bowen} to be 9.4 mb, while Carroll et
al. \cite{carroll} find 13 mb.  This latter gives an upper limit for $\Gamma$
of 3.6 MeV.

Our results are consistent with those obtained by Arndt, Strakovsky,
and Workman \cite{arndt}, who, using their partial wave analysis of
$K^+N$ data, exclude a $\Theta(1540)^+$ with a width of more than a
few MeV.  Our results are also consistent with, but more stringent
than those obtained by Nussinov \cite{shmuel} and by Haidenbauer and
Krein \cite{haidenbauer}.

\section{Comments}
A width of 1 MeV is quite uncommon for a hadronic decay.  For
comparison we consider the $\Lambda(1520)$, which decays by d wave to
${\overline K}N$ with a partial width of 7.2 MeV.  If the
$\Theta(1540)^+$
decays via p wave, it might be expected to be somewhat broader than
the
$\Lambda(1520)$.  Instead it is evidently much narrower.  

It is not possible to make quantitative statements of the same sort
using the photoproduction data reported by CLAS \cite{CLASI, CLASII}
or SAPHIR \cite{SAPHIR}.  However, qualitatively, the very small
apparent width suggests that nonresonant production cross sections
should be quite small, while the data of these experiments seem to
show quite visible effects.

The value for the width inferred from the DIANA and the limits derived
from the charge-exchange and total-cross-section measurements in
deuterium are not inconsistent.  However, they point to such a narrow
width that, if the $\Theta(1540)^+$ truly exists, it is exotic
dynamically as well as in its quantum numbers.

\section*{Acknowledgment}
This work was supported in part by the Director, Office of Science, Office of High Energy and Nuclear Physics, of the U.S. Department of Energy under Contract
DE-AC0376SF00098.


\begin{thebibliography}{99}
\bibitem{LEPS}T. Nakano et al. {\it Phys. Rev. Lett.} {\bf 91},012002 (2003).
\bibitem{DIANA}V. V. Barmin et al. {\it Phys. At. Nucl}{\bf 66}, 1715 (2003).
\bibitem{CLASI}S. Stepanyan et al., arXiv:hep-ex/0307018 (2003).
\bibitem{CLASII}V. Kubarovsky and S. Stepanyan,  arXiv:hep-ex/0307088 (2003).
\bibitem{SAPHIR}J. Barth et al.,{\it Phys. Lett.} {B572}, 127 (2003).
\bibitem{neutrino} A. E. Asratyan, A. G. Dogolenko, and
  M. A. Kubantsev, arXiv:hep-ex/0309042 (2003).
\bibitem{Diakonov}D. Diakonov, V. Petrov, and M. Polyakov, {\it
    Z. Phys.} {\bf A359}, 305 (1997).
\bibitem{slater}W. Slater et al., {\it Phys. Rev. Lett.} {\bf 7}, 378
  (1961).
\bibitem{damerell} C.J.S. Damerell et al., {\it Nucl. Phys.} {\bf
  B94}, 374 (1975).
\bibitem{hulthen}L. Hulth\'en, {\it Arkiv Mat. Astron. Fysik} {\bf
    28A}, No. 5 (1942).
\bibitem{bowen}T. Bowen et al., {\it Phys. Rev.} {\bf D}, 2599 (1970).
\bibitem{carroll}A. S. Carroll et al., {\it Phys. Lett.} {\bf 45B},
  531 (1973).
\bibitem{arndt}R. A. Arndt, I. I. Strakovsky, and R. L. Workman, {\it
    Phys. Rev. } {\bf C68}, 042201(R) (2003).
\bibitem{shmuel}S. Nussinov,  arXiv:hep-ph/0307357.
\bibitem{haidenbauer}J. Haidenbauer and G. Krein,  arXiv:hep-ph/0309243.
\end{thebibliography}
\end{document}